\RequirePackage[displaymath]{lineno} 
\documentclass[aps,prl,twocolumn,showpacs,amsmath,amssymb,nofootinbib]{revtex4-1}
\usepackage{epsfig}
\usepackage{graphicx}
\usepackage{dcolumn}
\usepackage{bm}
\usepackage{ltablex,booktabs}
\usepackage{overpic}
\usepackage{subfigure}
\usepackage{float}
\usepackage{color}

\usepackage{amsmath}
\usepackage{mathcomp}
\usepackage{mathrsfs}
\usepackage{multirow}
\usepackage{rotating}
\usepackage{amssymb}
\usepackage{gensymb}
\usepackage{amsmath}
\usepackage{tabularx}
\usepackage[bookmarksnumbered, pdfstartview=FitH,colorlinks,urlcolor=blue, citecolor=blue,linkcolor=blue] {hyperref}

\begin{document}
\normalsize
\parskip=5pt plus 1pt minus 1pt


\title{\boldmath Study of $\phi\to K\bar{K}$ and $K_{S}^{0}-K_{L}^{0}$ Asymmetry in the Amplitude Analysis of $D_{s}^{+}
  \to K_{S}^{0}K_{L}^{0}\pi^{+}$ Decays}

\author{
\begin{small}
\begin{center}
M.~Ablikim$^{1}$, M.~N.~Achasov$^{4,c}$, P.~Adlarson$^{77}$, X.~C.~Ai$^{82}$, R.~Aliberti$^{36}$, A.~Amoroso$^{76A,76C}$, Q.~An$^{73,59,a}$, Y.~Bai$^{58}$, O.~Bakina$^{37}$, Y.~Ban$^{47,h}$
% ...
\\
\vspace{0.2cm}
(BESIII Collaboration)\\
\vspace{0.2cm} {\it
$^{1}$ Institute of High Energy Physics, Beijing 100049, People's Republic of China\\
% ...
$^{4}$ Budker Institute of Nuclear Physics SB RAS (BINP), Novosibirsk 630090, Russia\\
% ...
$^{36}$ Johannes Gutenberg University of Mainz, Johann-Joachim-Becher-Weg 45, D-55099 Mainz, Germany\\
$^{37}$ Joint Institute for Nuclear Research, 141980 Dubna, Moscow region, Russia\\
% ...
$^{47}$ Peking University, Beijing 100871, People's Republic of China\\
% ...
$^{58}$ Southeast University, Nanjing 211100, People's Republic of China\\
$^{59}$ State Key Laboratory of Particle Detection and Electronics, Beijing 100049, Hefei 230026, People's Republic of China\\
% ...
$^{73}$ University of Science and Technology of China, Hefei 230026, People's Republic of China\\
% ...
$^{76}$ University of Turin and INFN, (A)University of Turin, I-10125, Turin, Italy; (B)University of Eastern Piedmont, I-15121, Alessandria, Italy; (C)INFN, I-10125, Turin, Italy\\
$^{77}$ Uppsala University, Box 516, SE-75120 Uppsala, Sweden\\
% ...
$^{82}$ Zhengzhou University, Zhengzhou 450001, People's Republic of China\\
\vspace{0.2cm}
$^{a}$ Deceased\\
% ...
$^{c}$ Also at the Novosibirsk State University, Novosibirsk, 630090, Russia\\
% ...
$^{h}$ Also at State Key Laboratory of Nuclear Physics and Technology, Peking University, Beijing 100871, People's Republic of China\\
% ...
}
\end{center}
\vspace{0.4cm}
\end{small}
}

\begin{abstract}
 Using $e^+e^-$ annihilation data corresponding to a total integrated
  luminosity of 7.33 $\rm fb^{-1}$ collected at center-of-mass
  energies between 4.128 and 4.226~GeV with the BESIII detector, we provide the first amplitude analysis and absolute branching fraction measurement of the hadronic decay $D_{s}^{+}
  \to K_{S}^{0}K_{L}^{0}\pi^{+}$. The branching fraction of $D_{s}^{+}
  \to K_{S}^{0}K_{L}^{0}\pi^{+}$ is determined to be $(1.86\pm0.06_{\rm stat}\pm0.03_{\rm syst})\%$.
  Combining the $\mathcal{B}[D_{s}^{+} \to \phi(\to K_{S}^0K_{L}^0) \pi^+]$ obtained in this work and the world average of $\mathcal{B}[D_{s}^{+} \to \phi(\to K^+K^-) \pi^+]$,
  we measure the relative branching fraction $\mathcal{B}(\phi \to K_S^0K_L^0)/\mathcal{B}(\phi \to K^+K^-)$=($0.593 \pm 0.023_{\rm stat} \pm 0.014_{\rm syst} \pm 0.016_{\phi\pi}$), where the third error is due to the uncertainty of the $\mathcal{B}(D_s^+ \to \phi \pi^+,\phi \to K^+K^-)$. Our result deviates from the Particle Data Group value by more than 3$\sigma$. 
Furthermore, the asymmetry of the branching fractions of $D^+_s\to K_{S}^0K^{*}(892)^{+}$ and $D^+_s\to K_{L}^0K^{*}(892)^{+}$,  $\frac{\mathcal{B}[D_{s}^{+} \to K_{S}^0K^{*}(892)^{+}]-\mathcal{B}[D_{s}^{+} \to K_{L}^0K^{*}(892)^{+}]}{\mathcal{B}[D_{s}^{+} \to K_{S}^0K^{*}(892)^{+}]+\mathcal{B}[D_{s}^{+} \to K_{L}^0K^{*}(892)^{+}]}$, is determined to be $(-14.5\pm 5.1_{\rm stat}\pm 1.8_{\rm syst})\%$.

\end{abstract}
\maketitle


The $\phi$ meson, with a large proportion 
of $s\bar{s}$ and a mass in the nonperturbative QCD region, 
is considered a potential carrier of 
the interaction between hadrons, as proposed in Hideki Yukawa's meson exchange theory~\cite{Hideki}. 
It is also a valuable probe for studying QCD matter formed in relativistic heavy-ion collisions~\cite{STAR}. 
Consequently, the precise measurement of its decay characteristics  
holds significant theoretical and experimental value for investigating the nonperturbative behavior of the strong interaction
and the properties of the nuclear force between baryons~\cite{PRC}, 
thus deepening understanding of the structure of hadronic matter. 

About 80\% of $\phi$ mesons decay into $\phi\to K\bar{K}$, and the relative branching fraction~(BF),  
$R_{\phi}\equiv\mathcal{B}(\phi \to K_S^0K_L^0)/\mathcal{B}(\phi \to K^+K^-)$, is naively expected to equal 1.0 due to isospin symmetry. Different theoretical predictions of $R_{\phi}$, however, fall within a relatively large range of 0.62-0.71~\cite{BRAMON,Flores, Fischbach,Benayoun}, taking into account phase-space difference, radiative corrections, isospin breaking etc. 
The experimental measurements of $R_{\phi}$ currently range from 0.64 to 0.89~\cite{PDG, OLYA, HBC72, HBC77, HBC78}. 
It is evident that although there is overlap between theoretical predictions and experimental measurements, further studies are required.

The Particle Data Group~(PDG) average value of $R_{\phi}=0.740 \pm 0.031$~\cite{PDG} has not been updated for many years~(while updated measurements of $R_{\phi}$ were reported by \textit{CMD2}~\cite{CMD2}, and \textit{CMD3}~\cite{CMD3} experiments in 1995 and 2018, respectively, they have not been taken into account for the averages by PDG). In addition, the measurements were primarily through $e^+e^-$ annihilation and $K - p$ scattering experiments~\cite{PDG, Parrour, Mattiuzzi, Dolinsky}, which usually suffer challenges from complex backgrounds and various interferences. Recently, Ref.~\cite{Dubn} used total cross-section measurements of the processes $e^+e^- \to K^+K^-/K_S^0K_L^0$, yielding an $R_{\phi}$ value of $0.644 \pm 0.017$ 
with the unitary and analytic model, suggesting the possibility that the experimental average may not be a reliable estimate of $R_{\phi}$ anymore. Therefore, exploring 
the reasons behind these differences and understanding the 
underlying mechanisms affecting $\phi$ meson decays requires a new and 
more accurate method to determine the BF of $\phi$ decays. The measurement of the BF of $D_s^+ \to \phi(\to K_S^0 K_L^0) \pi^+$, along with $D_s^+ \to \phi(\to K^+ K^-) \pi^+$~\cite{kkpi}, serves as a new approach to determine $R_{\phi}$ in a more controlled environment. In addition, the BESIII Collaboration found that the measured BF ratio $\mathcal{B}(\phi \to \pi^+ \pi^- \pi^0)/\mathcal{B}(\phi \to K^+K^-)$ deviates from the world average value by more than $4\sigma$ ~\cite{xiaoyu}. This further stimulates the urgent study of $\phi$ decays in the amplitude analysis of $D_s^+ \to K_S^0K_L^0\pi^+$ to explore the source of this tension. 

Moreover, the symmetry $\Gamma(\bar{K}^0)=2\Gamma(K_S^0)=2\Gamma(K_L^0)$ is usually used for the decay modes containing a neutral kaon $K_S^0$ or $K_L^0$. However, it is expected that the interference between Cabibbo-Favored and Doubly-Cabibbo-Suppressed transitions could
lead to a significant asymmetry, called $K_{S}^{0}-K_{L}^{0}$ asymmetry~\cite{Bigi, Rosner, Bhattacharya, Wang,kokp,cheng}. 
The $K_{S}^{0}-K_{L}^{0}$ asymmetry has not been observed in the $D \to K^0_{S,L}+{\rm Vector}$ system~\cite{omegaphi, harry, gao}. Phenomenological models, such as factorization-assisted topological amplitude\cite{Wang} and topological diagram approaches\cite{cheng}, predict non-zero $K_{S}^{0}-K_{L}^{0}$ asymmetry in the decay process of $D_s^+\to K^0_{S,L}+{\rm Vector}$. 
Measurements for the $K_{S}^{0}-K_{L}^{0}$ asymmetries serve as critical constraints on the dynamic models of charmed meson decays.
In the amplitude analysis of $D_s^+ \to K_S^0K_L^0\pi^+$, the asymmetry of $D_{s}^{+} \to K_{S}^{0}K^{*}(892)^{+}$
and $D_{s}^{+} \to K_{L}^{0}K^{*}(892)^{+}$ can be measured with most systematic uncertainties canceled. The obtained result will be crucial to better understand the dynamics of Cabibbo-Favored and Doubly-Cabibbo-Suppresse transitions. 



In this Letter, we present the first amplitude analysis and absolute BF measurement of the $D_{s}^{+} \to K_{S}^{0}K_{L}^{0}\pi^{+}$
decay using 7.33~$\rm fb^{-1}$ data samples collected at center-of-mass~(c.m.)
energies between 4.128-4.226~GeV with the BESIII
detector.  Charge conjugation is implied throughout this Letter.

The BESIII detector~\cite{Ablikim:2009aa, Ablikim:2019hff} records
symmetric $e^+e^-$ collisions provided by the BEPCII storage
ring~\cite{Yu:IPAC2016-TUYA01}. The cylindrical core of the BESIII
detector covers 93\% of the full solid angle and consists of a
helium-based multilayer drift chamber, a plastic scintillator
time-of-flight system, and a CsI(Tl) electromagnetic
calorimeter, which are all enclosed in a superconducting
solenoidal magnet providing a 1.0~T magnetic field.  The end cap time-of-flight
system was upgraded in 2015 using multigap resistive plate chamber
technology, providing a time resolution of 60 ps, which benefits 83\% of the data
used in this analysis~\cite{etof}.

Simulated data samples produced with {\small Geant4}-based~\cite{geant4} Monte
Carlo (MC) software, which includes the geometric description of the BESIII
detector and the detector response, are used to determine detection
efficiencies and to estimate backgrounds. The simulation models the beam energy
spread and initial state radiation in the $e^+e^-$ annihilations with the
generator {\sc kkmc}~\cite{ref:kkmc}. The inclusive MC sample includes the
production of open charm processes, the initial state radiation production of vector
charmonium(like) states, and the continuum processes. 
All particle decays are modeled with {\sc
evtgen}~\cite{ref:evtgen} using BFs 
either taken from the
PDG~\cite{PDG}, when available,
or otherwise estimated with {\sc lundcharm}~\cite{ref:lundcharm}.
Final-state radiation from charged 
final-state particles is incorporated using the {\sc photos} package~\cite{photos2}.

The process
$e^{+}e^{-} \to D_{s}^{*\pm}D_{s}^{\mp}\to \gamma D_{s}^{+}D_{s}^{-}$
allows studies of $D_s^{+}$ decays using a tag technique~\cite{MarkIII-tag, Ke:2023qzc}.
Two types of samples are used: single tag~(ST) and
double tag~(DT). 
An ST candidate is an event where only the $D_{s}^{-}$ meson is reconstructed through particular hadronic decays, designated as ``tag'', is reconstructed through one of 
ten hadronic decay modes: $K_{S}^{0}K^{-}$, $K^{+}K^{-}\pi^{-}$,
$K_{S}^{0}K^{-}\pi^{0}$, $K^{+}K^{-}\pi^{-}\pi^{0}$, $K_{S}^{0}K^{-}\pi^{-}\pi^{+}$,
$K_{S}^{0}K^{+}\pi^{-}\pi^{-}$, $\pi^{-}\pi^{-}\pi^{+}$, $\pi^{-}\eta$, $\pi^{-}\eta^{\prime}$,
and $K^{-}\pi^{-}\pi^{+}$. A DT candidate is an event where the $D_{s}^{+}$ is reconstructed through $D_{s}^{+} \to K^0_{S}K^0_{L}\pi^{+}$ in addition to the $D_{s}^{-}$ being reconstructed through the tag modes. A detailed description of selection conditions
concerning charged and neutral particle candidates, the mass recoiling against
$D_s^{\pm}$ candidates, and the mass of the tag candidates is provided in
Refs.~\cite{ref:a0980, ref:Kspipi0, ref:KsKpi0, ref:KsKspi,
ref:KsKpipi}.


After a $D_s^-$ tag candidate is identified, we reconstruct the signal $D_{s}^{+} \to K^0_{S}K^0_{L}\pi^{+}$ candidate recoiling against the
tag by requiring two positively charged particles identified as $\pi^+$, one negatively charged $\pi^-$,
and at least one more photon to reconstruct the transition photon of
$D_{s}^{*\pm} \to \gamma D_{s}^{\pm}$. 
The four-momentum of the $K_L^0$ needs to be determined, which is calculated with the momentum of the initial $e^+e^-$ system and other detected particles. 
If there are multiple 
signal candidates for $K_S^0$, the best candidate with the minimum $\chi^2$ of a four-constraint~(4C) kinematic fit is chosen.
The total four-momentum is constrained to the four-momentum of the initial $e^+e^-$ beams. The invariant masses of the tag $D_s^-$, the signal $D_s^+$, the $D_s^*$, and $K_S^0$ are constrained to their PDG values~\cite{PDG}. 
The two cases $D_s^{*+} \to D_s^{+}\gamma$ and $D_s^{*-} \to D_s^{-}\gamma$ are considered.
The combination with the minimum $\chi^2$ is chosen.
Furthermore, the square of the recoil mass against the transition
photon and the tag $D_s^-(M_{\rm rec}^{\prime 2})$ is expected to peak at the known $D_s^{\pm}$ meson mass squared before the kinematic fit for signal $D_s^{*\pm}D_s^{\mp}$ events, and must satisfy 3.80  ~$<M^{\prime 2}_{\rm rec}<$~4.0  GeV$^{2}$/$c^4$.
The requirement that the number of additional $\pi^0$($N_{\pi^0}$) composed of unused photons in the ST candidate selection and $D_s^{*\pm} \to \gamma D_s^{\pm}$ is equal to zero is applied to suppress backgrounds. 
Here, since $K_L^0$ may cause fake photons in the electromagnetic calorimeter, the angles of any 
photons that form $\pi^0$s and the shower produced by $K^0_L$ are  required to be greater than 10°.
 

The purity is determined by fitting the missing mass squared ($M_{\rm miss}^2$) of the $K_L^0$ candidates after the 4C kinematic fit, which is 
defined as
\begin{equation}
  M_{\rm miss}^{2}=\frac{1}{c^2}(\vec{p}_{\rm c.m.}-\vec{p}_{\rm tag}-\vec{p}_{K_S^0}-\vec{p}_{\pi^+}-\vec{p}_{\gamma})^2, 
  \label{mm2}
\end{equation}
where $\vec{p}_{\rm c.m.}$ is the four-momentum of the $e^+e^-$ c.m. system,
$\vec{p}_{\rm tag}$ and $\vec{p}_{i}$ ($i=K_S^0, \pi^+, \gamma$) are the four-momenta of the tag candidate and the final-state
particle $i$ on the signal side. The $M_{\rm miss}^{2}$ mass 
window, [0.21, 0.29]~GeV$^{2}$/$c^4$, is applied on the signal candidates for the amplitude analysis. 
Signal candidates from all center-of-mass energy points  are combined into a single sample, resulting in 2310 events with a purity of $f_s =
(78.2\pm 1.0)$\%. Here, the peaking background from $D_s^+\to K_S^0K_S^0\pi^+$ is 4.3\%, and simulated based on the amplitude analysis by BESIII ~\cite{ref:KsKspi}.

Based on the 4C kinematic fit, an additional constraint on the mass of $K_L^0$ is added (five-constraint) and the four-momenta of the final-state particles of the five-constraint kinematic fit are used for the amplitude analysis to ensure that all candidates fall within the phase-space boundary.
The isobar formulation is used in the covariant tensor formalism~\cite{covariant-tensors}.
The total signal amplitude $\mathcal{M}=\begin{matrix}\sum \rho_{n}e^{i\varphi_{n}}A_{n}\end{matrix}$, is described by a coherent sum of the amplitudes from all intermediate processes, where
$n$ represents the $n$th intermediate state with magnitude $\rho_{n}$ and phase $\varphi_{n}$.
The decay amplitude $A_{n}$ depends on the phase space and intermediate resonance spin, and is given by $A_{n} = P_{n}S_{n}F_{n}^{r}F_{n}^{D}$, where $S_{n}$ and
$F_{n}^{r(D)}$ are the spin factor~\cite{covariant-tensors} and the
Blatt-Weisskopf barrier factor of the intermediate state (the $D_{s}^{\pm}$
meson)~\cite{Blatt}, respectively, and $P_{n}$ is the propagator of the intermediate resonance, which is the relativistic Breit-Wigner amplitude~\cite{RBW}.

The unbinned maximum likelihood method is adopted in the amplitude analysis.
A combined probability density
function~(PDF) for the signal and background hypotheses is constructed, with the four-momenta of the final-state particles. 
The signal PDF is constructed from the total
amplitude $\mathcal{M}$. The background PDF, $B$, is constructed from a background shape
derived from the inclusive MC samples using the XGBoost package~\cite{xgboost1,xgboost2}.
This background PDF is then added to the
signal PDF incoherently. The likelihood function is written as
\begin{eqnarray}
  \begin{aligned}
             \mathcal{L}=\prod_{j} &\left[\frac{\epsilon f_s\left|\mathcal{M}\right|^{2}\,R_3}{\int \epsilon\left|\mathcal{M}\right|^{2}\,R_{3}dp_j}
          +\frac{(1-f_s)BR_{3}}{\int  B\,R_{3}dp_j}\right], 
  \end{aligned}
\end{eqnarray}
where $j$ runs over the selected events, $p_j^{\mu}$ represents the four-momenta of the final-state particles, and $\epsilon$ is the detection efficiency determined with a MC sample of $D_s^+\to K_S^0K_L^0\pi^+$ uniformly distributed over the Dalitz plot. 
Finally, $f_s$ and $R_{3} dp_j$ denote the purity and three-body phase-space element, respectively. Among them, $\mathcal{M}$, $B$, $\epsilon$, and $R_{3}$ are all functions of the variable $p_j^{\mu}$.
The normalization integral
in the denominator is determined by an MC technique as described in
Refs.~\cite{ref:Kspipi0, ref:KsKpipi, ref:KsKspi, ref:kspieta, ref:KsKpi0}.

\begin{figure}[!htp]
  \centering
  \includegraphics[width=0.235\textwidth]{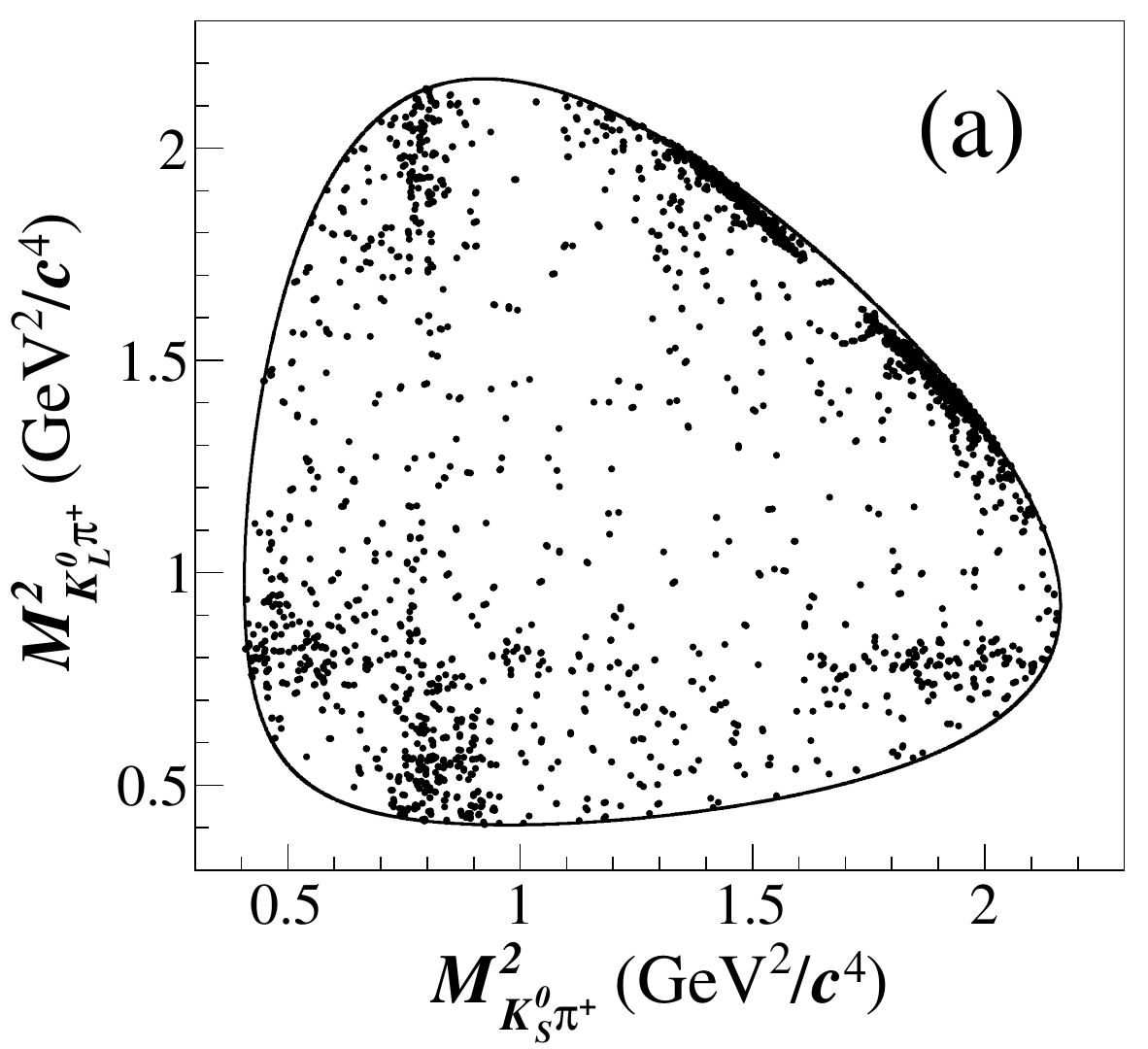}
  \includegraphics[width=0.235\textwidth]{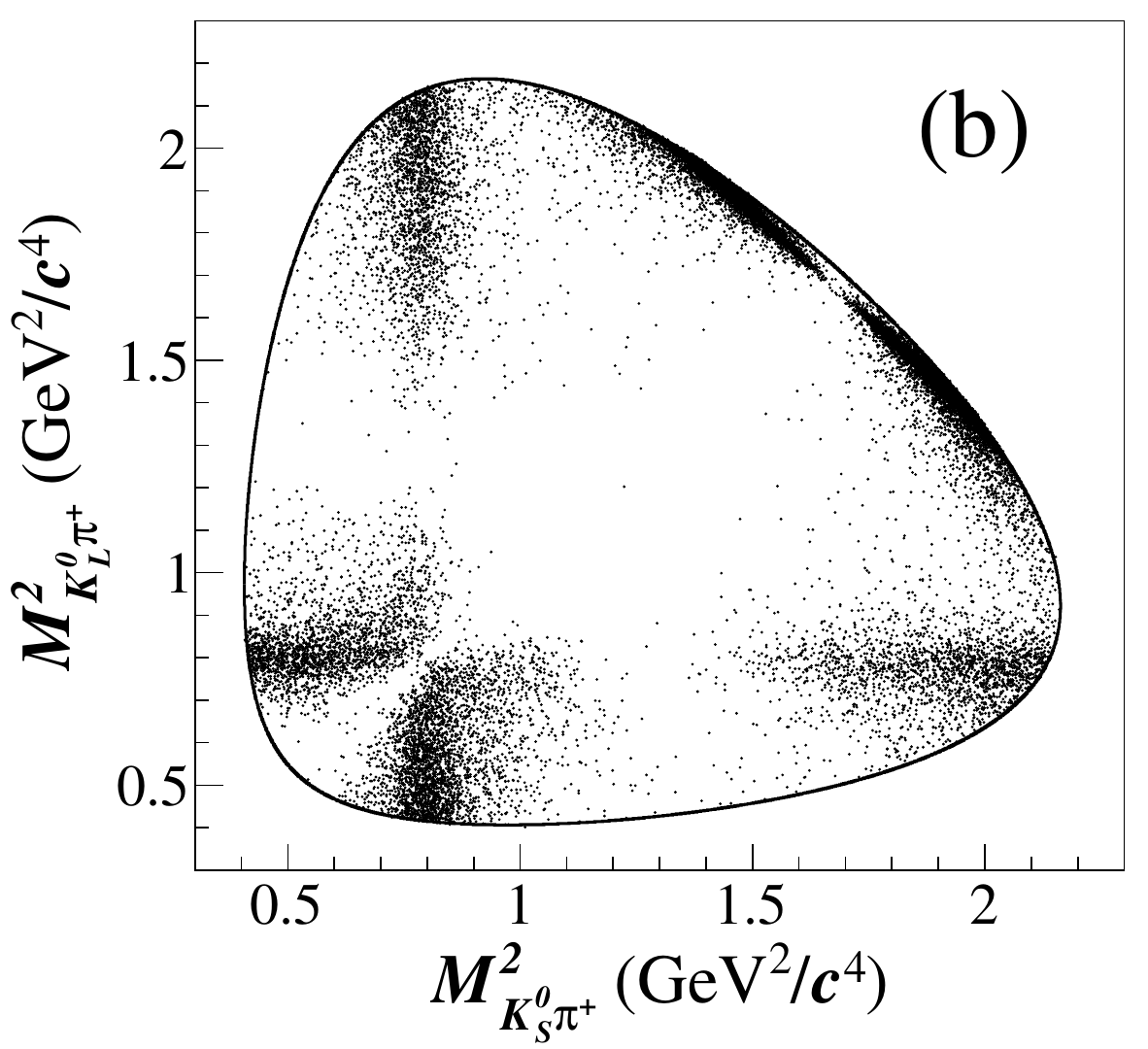}
    \caption{Dalitz plots of $M^{2}_{K_L^0\pi^+}$ versus~$M^{2}_{K_S^0\pi^+}$
      for $D^+_s\to K^0_SK^0_L\pi^+$, of (a) the sum of all data samples and (b) the signal MC samples generated
      based on the amplitude analysis. The black curve indicates the
      kinematic boundary.}
    \label{fig:dalitz}
\end{figure}

\begin{figure}[!htbp]
  \centering
  \includegraphics[width=0.235\textwidth]{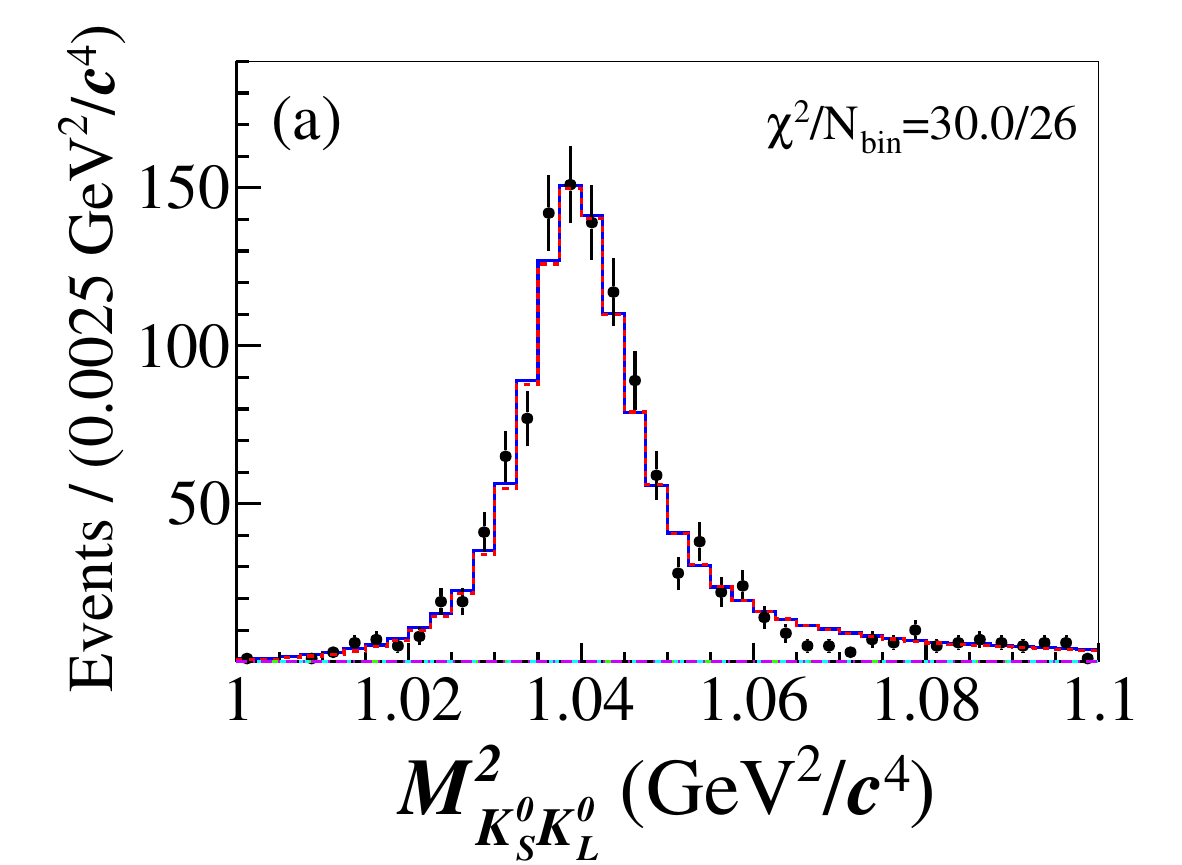}
  \includegraphics[width=0.235\textwidth]{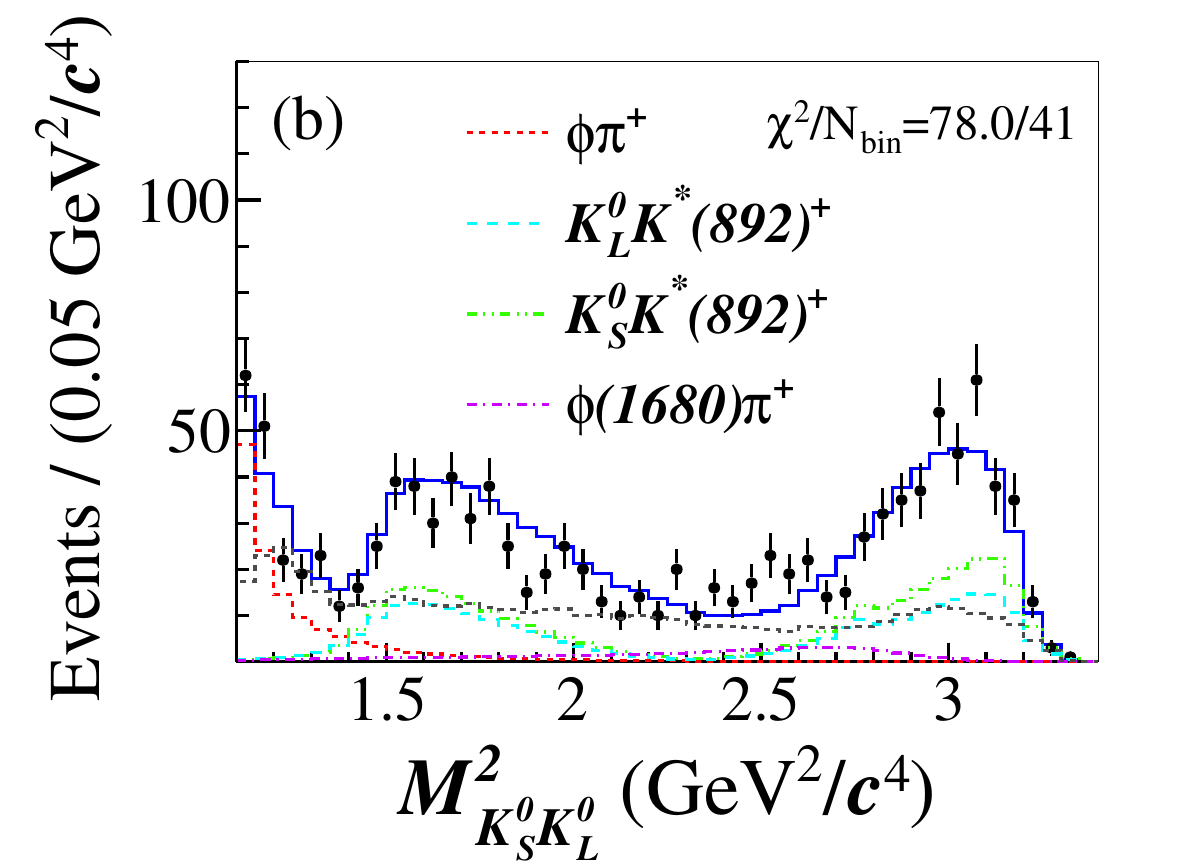}
  \includegraphics[width=0.235\textwidth]{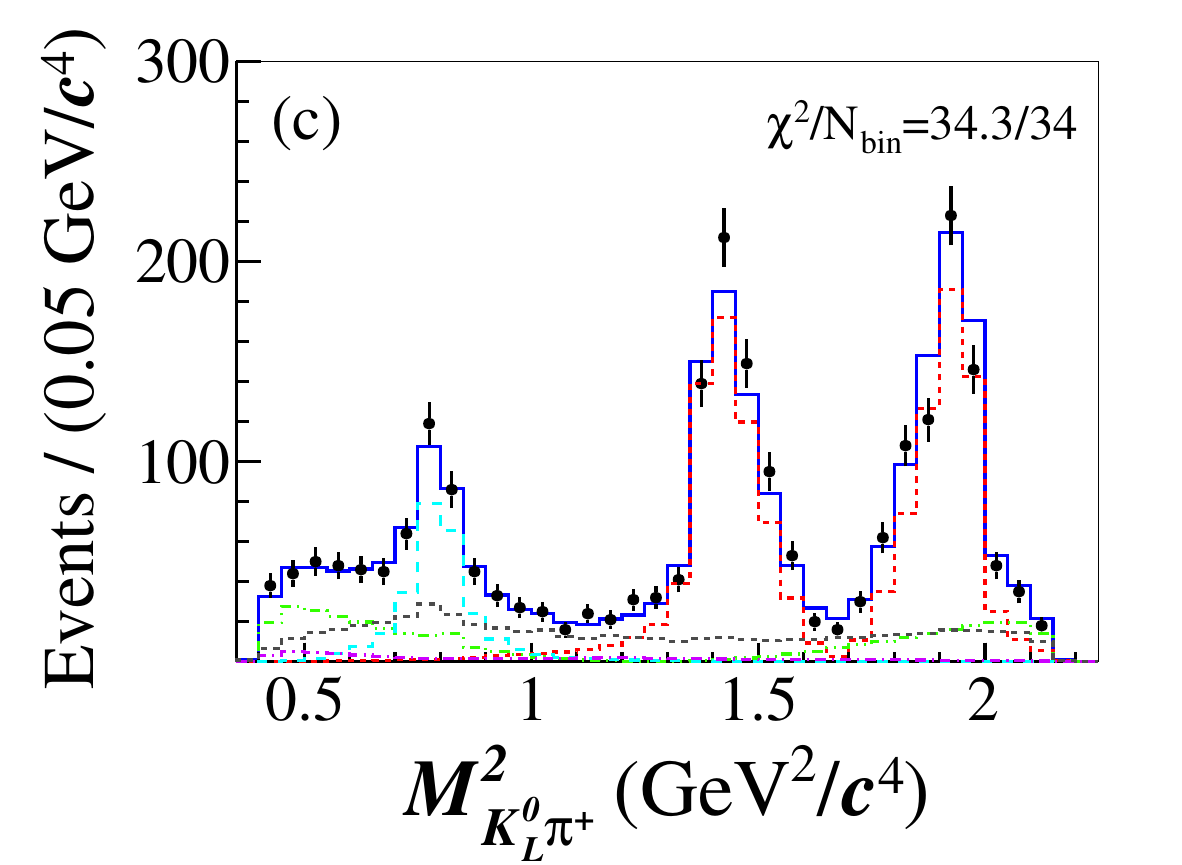}
  \includegraphics[width=0.235\textwidth]{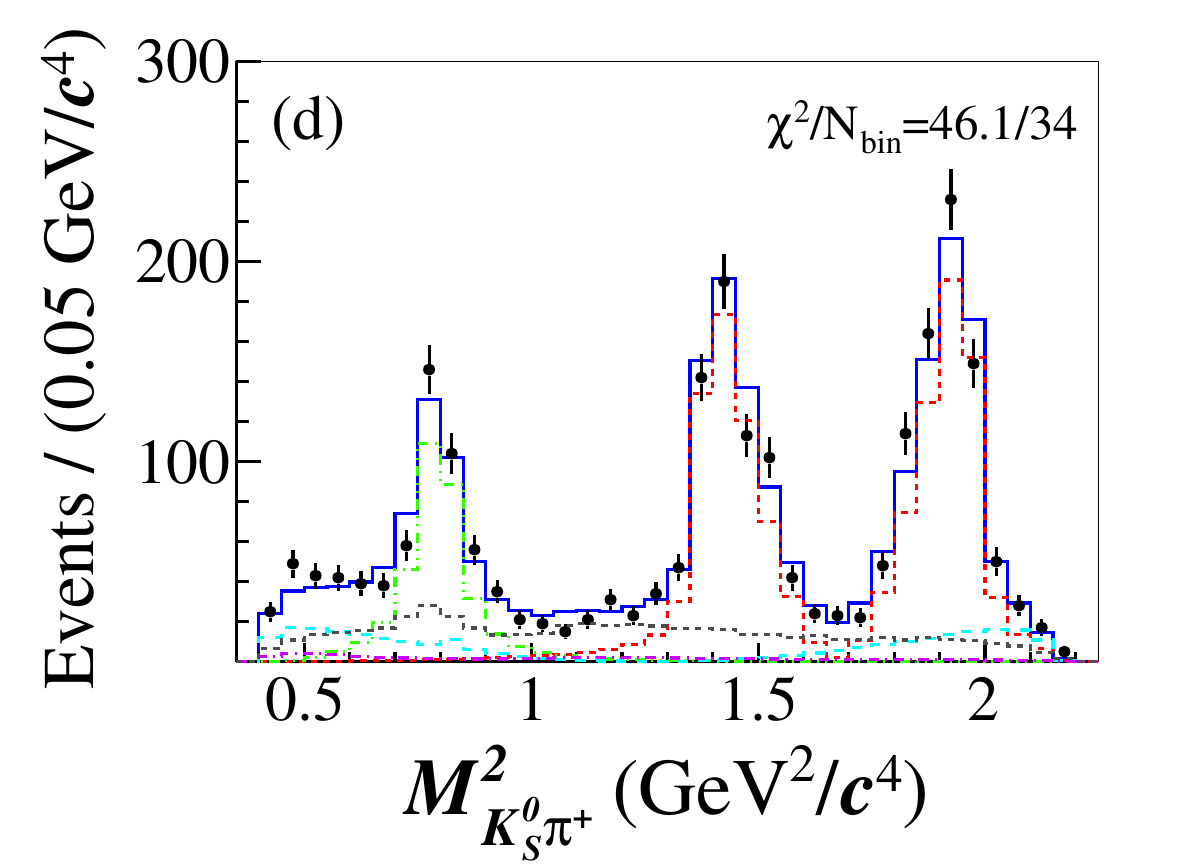}
  \caption{
    Distributions of (a) $M_{K_S^0K_L^0}^{2}$ less than 1.1~GeV$^{2}$/$c^4$, (b) $M_{K_S^0K_L^0}^{2}$ greater than 1.1~GeV$^{2}$/$c^4$, (c) $M_{K_L^0\pi^+}^{2}$ and (d) $M_{K_S^0\pi^+}^{2}$ from the nominal fit. The data samples
    are represented by points with error bars, the fit results by blue 
    lines, and the backgrounds by gray dashed lines. Colored dashed lines show the individual components of the fit model. The numbers in the upper right corner represent the  goodness of fit, $\chi^2$/$N_{\rm bin}$, where $N_{\rm bin}$ denotes the number of bins for each projection. In the calculation, adjacent bins are merged to ensure that each bin has no fewer than 10 entries.
    }

  \label{dalitz-projection}
\end{figure}

The Dalitz plot of $M^2_{K_L^0\pi^+}$ versus $M^2_{K_S^0\pi^+}$ from all data samples is shown in
Fig.~\ref{fig:dalitz}(a). The diagonal band in the upper right corner is caused by the process $D_s^+\to \phi \pi^+$, the vertical and horizontal bands around
0.8~GeV$^2$/$c^4$ are $D_s^+\to K_L^0K^{*}(892)^+$ and $D_s^+\to K_S^0K^{*}(892)^+$, respectively.
The $D_s^+\to \phi \pi^+$ process is used as a reference so that the magnitudes and phases of other amplitudes can be fitted as relative values to this reference amplitude. The purity is fixed in the fit.
Other possible contributing resonances such as $K^{*}(1410)^{+}$,
$K^{*}_{0}(1430)^{+}$, $K^{*}_{2}(1430)^{+}$, $(K_S^0\pi^{+})_{\mathcal{S}-\rm wave}$, $(K_L^0\pi^{+})_{\mathcal{S}-\rm wave}$, $\phi(1680)$, $\rho(980)$, $\rho(1450)$, $\rho(1700)$, $\omega(1650)$, $\omega_3(1670)$, and $\rho_3(1690)$ are added to the
fit one at a time.
The masses and widths of all resonances
are fixed to their PDG values~\cite{PDG}.
The statistical significance of each new amplitude
is calculated from the change of the log-likelihood taking the change in the number of
degrees of freedom into account. Various combinations of these resonances are also tested. 
Only the amplitudes $D_s^+\to \phi \pi^+$, $D_s^+\to K_L^0K^{*}(892)^+$, $D_s^+\to K_S^0K^{*}(892)^+$, and $D_s^+\to \phi(1680) \pi^+$ are found, and no other contribution has a significance greater than $3\sigma$.
The Dalitz plot of the signal MC sample
generated based on the result of the amplitude analysis is shown in Fig.~\ref{fig:dalitz}(b).
The mass projections of the fit are shown in Fig.~\ref{dalitz-projection}.

The contribution of the $n$th amplitude relative to the total BF is quantified
by the fit fraction~(FF) defined as
${\rm FF}_{n} = \int \left|\rho_{n}A_{n}\right|^{2}R_3 dp_{j}/\int\left|\mathcal{M}\right|^{2}R_3dp_{j}$.
The FFs of both amplitudes and the phase differences relative to the reference process are listed in Table~\ref{fit-result}. The phase difference of nearly $\pi$ radians between $D_{s}^{+} \to K_{L}^0K^{*}(892)^{+}$ and $D_{s}^{+} \to K_{L}^0K^{*}(892)^{+}$ decays leads to destructive interference between them. The sum of the three FFs is 107.2$\%$.
The interference fit fraction between amplitudes and the correlation matrix of amplitude parameters are given in Supplemental Material~\cite{SuppM}.
The asymmetry of the branching fractions of $D^+_s\to K_{S}^0K^{*}(892)^{+}$ and $D^+_s\to K_{L}^0K^{*}(892)^{+}$ is determined to be $\frac{\mathcal{B}[D_{s}^{+} \to K_{S}^0K^{*}(892)^{+}]-\mathcal{B}[D_{s}^{+} \to K_{L}^0K^{*}(892)^{+}]}{\mathcal{B}[D_{s}^{+} \to K_{S}^0K^{*}(892)^{+}]+\mathcal{B}[D_{s}^{+} \to K_{L}^0K^{*}(892)^{+}]}$=$(-14.5 \pm 5.1_{\rm stat}\pm 1.8_{\rm syst})\%$, where the correlation of uncertainties between $D_{s}^{+} \to K_{S}^0K^{*}(892)^{+}$ and $D_{s}^{+} \to K_{L}^0K^{*}(892)^{+}$ is considered. 

\begin{table*}[!htbp]
  \caption{Phases, FFs, BFs, and statistical significances ($\sigma$) of intermediate processes in $D_s^+\to K_S^0K_L^0\pi^+$. 
  The first and second
    uncertainties are statistical and systematic, respectively. 
    }
  \label{fit-result}
  \begin{center}
    \begin{tabular}{lccccc}
      \hline \hline
            Amplitude                                  & Phase~(rad)                     & FF~(\%)     & BF~(\%)                &$\sigma$\\
      \hline
      $D_{s}^{+} \to \phi \pi^{+}$        & 0.0(fixed)                     & $70.2 \pm 1.4 \pm 1.2$ & $1.31 \pm 0.05 \pm 0.03$ &$>$10 \\ 
      $D_{s}^{+} \to K_{L}^{0}K^{*}(892)^{+}$         &\phantom{0}$1.10 \pm 0.20 \pm 0.20$ & $19.3\pm 1.2  \pm 1.0$  & $0.36 \pm 0.03 \pm 0.02$&$>$10 \\  
      $D_{s}^{+} \to K_{S}^{0}K^{*}(892)^{+}$         & $-1.93 \pm 0.20 \pm 0.25$ & $14.4 \pm 1.2 \pm 1.2$ & $0.27 \pm 0.02 \pm 0.02$ &$>$10 \\
           $D_{s}^{+} \to \phi(1680) \pi^{+}$        & $-1.26 \pm 0.22 \pm 0.14$                     & \phantom{0}$3.3 \pm 0.9 \pm 0.7$ & $0.06 \pm 0.02 \pm 0.02$ &$4.3$ \\  
      
      \hline \hline
    \end{tabular}
  \end{center}
\end{table*}

The systematic uncertainties related to the amplitude analysis, including the phase
difference, FFs and $K_S^0-K_L^0$ asymmetry, are determined by the differences between
the results of the nominal fit and the alternative fits. The masses and widths of the $\phi$, $K^{*}(892)^{+}$, and $\phi(1680)$ are
shifted by their uncertainties~\cite{PDG}. The radii of the Blatt-Weisskopf
barrier factors are varied from their nominal values of $5$ and $3$~GeV$^{-1}$ (for the $D_s^+$ meson
and the intermediate resonances, respectively)  by $\pm 1$~GeV$^{-1}$. 
The uncertainties associated with the
size of the background sample are studied by varying the purity within its statistical
uncertainty. An alternative background sample is used to determine the background PDF, where
the relative fractions of background processes from direct $q\bar{q}$ and
non-$D_{s}^{*\pm}D_{s}^{\mp}$ open-charm processes are varied by the statistical uncertainties of the known cross sections.
The uncertainty from the peaking background is also considered based on the uncertainty from the measurement of $D_{s}^{+} \to K_{S}^{0}K_{S}^{0}\pi^{+}$~\cite{PDG} with one $K_{S}^{0} \to \pi^{0}\pi^{0}$.
In addition, potential intermediate resonances that were excluded from the baseline fit model due to low significance are introduced individually as systematic variations of the model. The following resonances are considered: $K^{*}(1410)^{+}$,
$K^{*}_{0}(1430)^{+}$, $K^{*}_{2}(1430)^{+}$, $(K_S^0\pi^{+})_{\mathcal{S}-\rm wave}$, $(K_L^0\pi^{+})_{\mathcal{S}-\rm wave}$, $\rho(1450)$, $\rho(1700)$, $\omega(1650)$, $\omega_3(1670)$, and $\rho_3(1690)$.
The total uncertainties are obtained by adding these contributions in quadrature. A breakdown of the systematic uncertainty by source can be found in Supplemental Material~\cite{SuppM}. 
Correlations between the systematic uncertainties on the $D^+_s\to K_{S/L}^0K^{*}(892)^{+}$ branching fractions are accounted for when calculating the asymmetry.

\begin{figure}[htp]
  \begin{center}
    \includegraphics[width=0.40\textwidth]{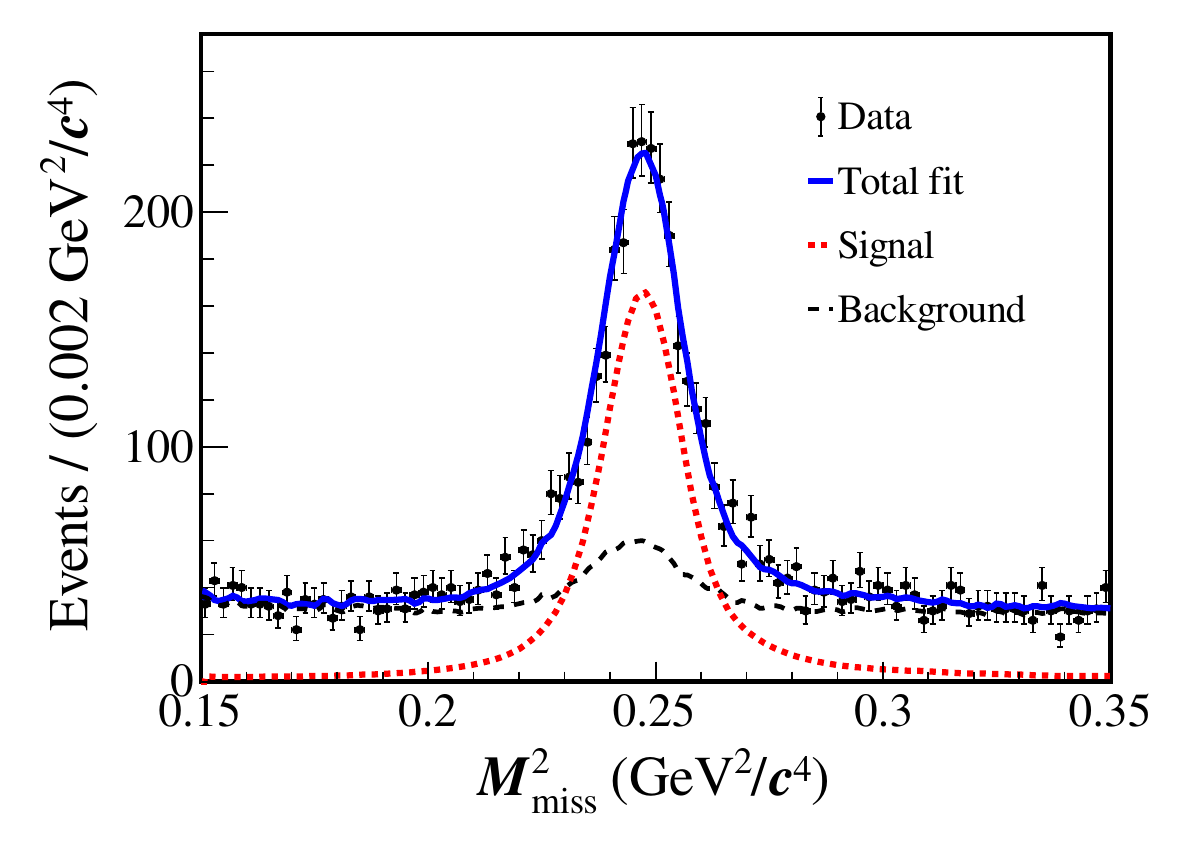}
    \caption{Fit to the $M_{\rm miss}^{2}$ distribution of the DT
      signal candidates. The data samples are represented by points with error bars, the signal contributions by the red dashed line, the total fit results by the solid blue line,  
      and the background
      contributions by the dashed black line. 
      }
    \label{fig:DT_fit}
  \end{center}
\end{figure}

The selection criteria for the BF measurement are the same as for the amplitude analysis with the exception that the requirements of $N_{\pi^0}=0$ and 0.21~$<$$M_{\rm miss}^2$$<$~0.29 GeV$^{2}$/$c^4$ are removed.
This is done to maximize the statistical significance of the signal and minimize the systematic uncertainties as much as possible.
 The BF is given by~\cite{ref:Kspipi0, ref:KsKpipi}
\begin{eqnarray} \begin{aligned}
    \mathcal{B}_{\text{sig}}=\frac{N_{\text{total,sig}}^{\text{DT}}}{\mathcal{B}_{\rm sub}\sum_{\alpha,   
        i} N_{\alpha, i}^{\text{ST}}\epsilon^{\text{DT}}_{\alpha,\text{sig},
        i}/\epsilon_{\alpha, i}^{\text{ST}}},\, \label{eq:Bsig-gen}
\end{aligned} \end{eqnarray} 
where $\alpha$ runs over the various tag modes, and $i$ denotes the
different c.m. energies, and $\mathcal{B}_{\rm sub}$ represent the product of the BFs of $D_s^* \to \gamma D_s$ and $K_S^0 \to \pi^+ \pi^-$. The ST yields in data $N_{\alpha,
  i}^{\text{ST}}$ and the DT yield $N_{\text{total,sig}}^{\text{DT}}$
are determined by fitting the mass of $D_s^-$ and $M_{\rm miss}^2$
distributions, respectively. The fit to the $M_{\rm miss}^2$ distribution is shown in 
Fig.~\ref{fig:DT_fit}. The signal shape is modeled
with the MC-simulated shape convolved with a Gaussian function. 
The dominant peaking background is $D_s^+\to K_S^0K_S^0\pi^+$ with one $K_S^0\to \pi^0\pi^0$. 
This peaking background is modeled by the MC-simulated shape based on the amplitude analysis~\cite{ref:KsKspi}, with a size Gaussian-constrained to the expected yield $584\pm33$ according to its measured BF ~\cite{PDG}. 
The shape of combinatorial background is derived from the inclusive MC samples and its size is floated in the fit. The corresponding efficiencies
$\epsilon$ are obtained by analyzing the inclusive MC samples, with
the signal MC events of $D_s^+\to K_S^0K_L^0\pi^+$ generated based on
the results of the amplitude analysis. 
The details of ST yields, ST efficiencies, and DT efficiencies can be found in Supplemental Material~\cite{SuppM}.
The total ST yields of all tag
modes and the DT yields are $665265\pm2750$ and $2349\pm76$,
respectively. The BF of $D_s^+\to K_S^0K_L^0\pi^+$ is determined to be
$(1.86\pm0.06_{\rm stat}\pm0.03_{\rm syst})\%$.

The following systematic uncertainties are considered in the BF measurement.
The relative uncertainty in the total number of ST $D_s^-$ mesons is assigned
to be 0.4\%.
The uncertainty related to a nonpeaking background shape in the fit to a $M_{\rm miss}^2$ 
distribution is assigned by repeating the fit with the MC background components varied by $\pm 30\%$.
The $\pi^+$ particle identification and tracking efficiencies 
are studied with $e^+e^- \to K^+K^-\pi^+\pi^-$ and the corresponding uncertainties are assigned to be 0.5\% and 0.2\%, respectively. 
The uncertainty for the $K_{S}^{0}$ reconstruction is 1.0\% using control samples of $J/\psi\to K_{S}^{0}K^{\pm}\pi^{\mp}$ and $J/\psi \to \phi K_{S}^{0}K^{\pm}\pi^{\mp}$.
The systematic uncertainty of photon reconstruction is assigned as 1.0\% with the control sample of $D_s^+\to K_S^0K^+$.
The uncertainty from the signal MC model based on
the results of the amplitude analysis is studied by varying the fit
parameters according to the covariance matrix. The change of signal
efficiency, 0.1\%, is assigned as the uncertainty.  
The uncertainties from the quoted BFs of $K_S^0 \rightarrow \pi^+ \pi^-$ and $D_{s}^{*\pm} \to \gamma D_{s}^{\pm}$ are 0.1\% and 0.4\%~\cite{PDG}, respectively. The uncertainty due to the limited signal MC sample size is 0.3\%. 
The total uncertainty is determined by adding all the contributions in quadrature and is 1.6\%.

\begin{figure}[htp]
  \begin{center}
    \includegraphics[width=0.45\textwidth]{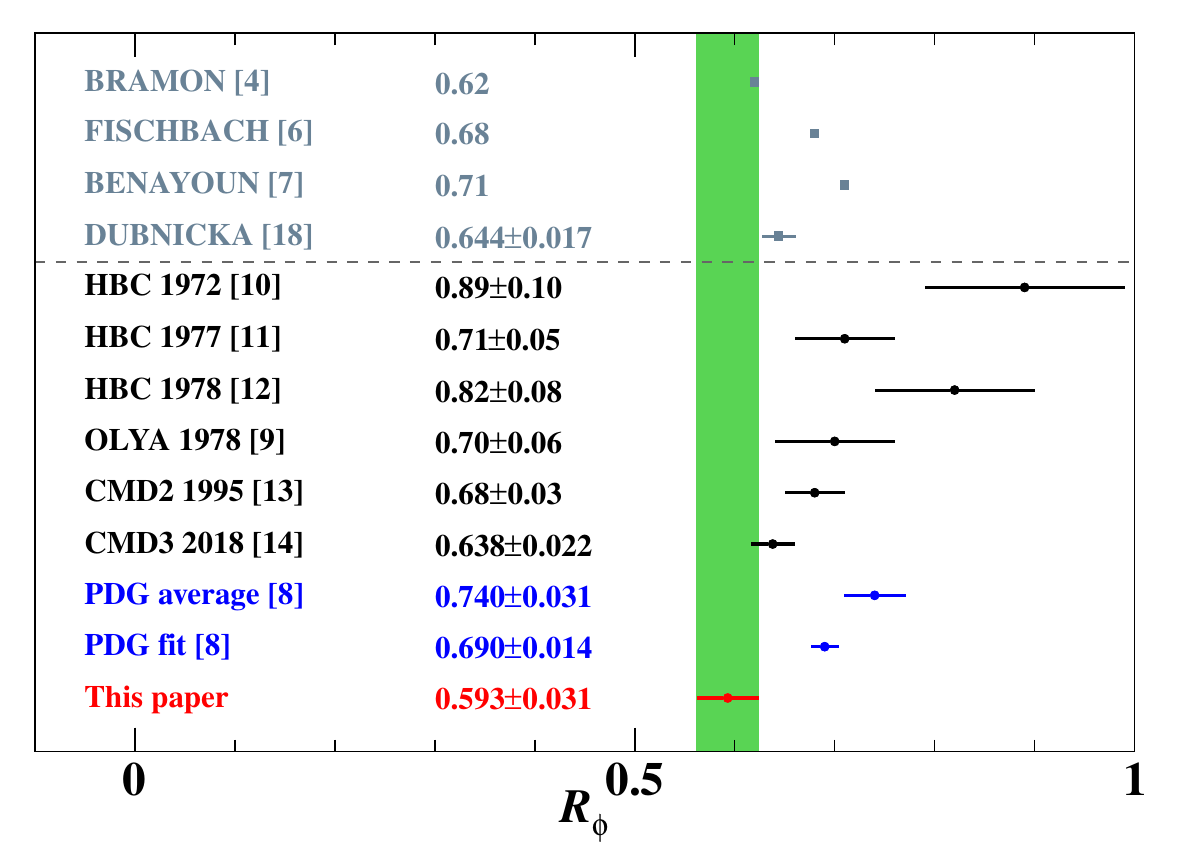}
    \caption{Comparison of the results for $R_{\phi}$ measured in this analysis and the HBC, OLYA, CMD2, and CMD3 experiments. Above the dotted line are the theoretical calculations, below are the experimental results. The green band present the total uncertainty obtained in this work.
      }
    \label{fig:comp}
  \end{center}
\end{figure}

In summary, we have presented the first amplitude analysis and BF measurement of the hadronic decay $D_{s}^{+} \to K_{S}^0K_{L}^0\pi^{+}$ using 7.33 fb$^{-1}$ of $e^+e^-$
annihilation data taken at c.m. energies between 4.128 and
4.226~GeV. The amplitude analysis results are listed in Table~\ref{fit-result}. With a detection efficiency obtained based on the amplitude analysis model, we obtain ${\cal B}(D_{s}^{+} \to K_{S}^0K_{L}^0\pi^{+})=(1.86\pm0.06_{\rm stat}\pm0.03_{\rm syst})\%$. The BFs of intermediate processes are calculated via ${\cal B}_i={\rm FF}_i\times{\cal B}(D_{s}^{+} \to K_{S}^0K_{L}^0\pi^{+})$ and ${\cal B}(D_{s}^{+} \to \phi \pi^{+}, \phi \to K_{S}^0K_{L}^0)$ is determined to be $(1.31\pm 0.05_{\rm stat.} \pm 0.03_{\rm syst.})\%$. With the PDG value of ${\cal B}(D_s^+ \to \phi \pi^+,\phi \to K^+K^-)=(2.21\pm0.06)\%$~\cite{PDG}, we determine a relative BF between $\phi \to K_{S}^0K_{L}^0$ and $\phi \to K^+K^-$ to be $R_{\phi}$=($0.593 \pm 0.023_{\rm stat} \pm 0.014_{\rm syst} \pm 0.016_{\rm \phi \pi}$), where the third error is due to the uncertainty of the PDG value of $\mathcal{B}(D_s^+ \to \phi \pi^+,\phi \to K^+K^-)$~\cite{PDG}. The obtained $R_{\phi}$ is consistent with theoretical expectations as reported in Ref.~\cite{BRAMON}. However, it is $(1.0-2.8)\sigma$ below all previous measurements, see Fig.~\ref{fig:comp}, and deviates from the PDG average (PDG fit) by $3.2\sigma$ ($2.6\sigma$). Note that the earlier measurement of $\mathcal{B}(\phi \to \pi^+ \pi^- \pi^0)/\mathcal{B}(\phi \to K^+K^-)$ by BESIII~\cite{xiaoyu} also
significantly deviates from the PDG values that were obtained in $e^+e^-$ annihilation and $K-p$ scattering experiments. 
To further explore the reasons behind these differences and to understand the underlying mechanisms that influence the BFs of $\phi$ meson decays, more precise measurements are needed in the future.

In addition, the $K_S^0-K_L^0$ asymmetry in $D_s^+ \to \bar{K}^0K^{*}(892)^+$ is determined to be $\frac{\mathcal{B}[D_{s}^{+} \to K_{S}^0K^{*}(892)^{+}]-\mathcal{B}[D_{s}^{+} \to K_{L}^0K^{*}(892)^{+}]}{\mathcal{B}[D_{s}^{+} \to K_{S}^0K^{*}(892)^{+}]+\mathcal{B}[D_{s}^{+} \to K_{L}^0K^{*}(892)^{+}]}=(-14.5 \pm 5.1_{\rm stat}\pm 1.8_{\rm syst})\%$. The predicted $K_S^0-K_L^0$ asymmetries from different approaches, as well as the measured value, are summarized in Table~\ref{kskl}. This is the first observation of the $K_S^0-K_L^0$ asymmetry in the $D \to K^0_{S,L}+{\rm Vector}$ system of charmed meson decays. 
\begin{table*}[!htbp]
  \caption{Predictions for $K_S^0-K_L^0$ asymmetries in $D_s^+ \to \bar{K}^0K^{*+}$ decays from different phenomenological models and our measurement result(FAT, factorization-assisted topological amplitude model; TDA, topological diagram approach). The unit is \%.
    }
  \label{kskl}
  \begin{center}
    \begin{tabular}{lccccc}
      \hline \hline
            Model                                  & TDA(F4)                     & TDA(F$1'$)     & FAT               &This work\\
      \hline
      $D_s^+ \to \bar{K}^0K^{*+}$        & $-16.4 \pm 3.2$                  & $-15.9 \pm 2.8$ &$-7.0  \pm 3.2$
      &$-14.5 \pm 5.1 \pm 1.8$  \\ 
      
      \hline \hline
    \end{tabular}
  \end{center}
\end{table*}

The BESIII Collaboration thanks the staff of BEPCII~(https://cstr.cn/31109.02.BEPC) and the IHEP computing center for their strong support. This work is supported in part by National Key R\&D Program of China under Contracts Nos. 2023YFA1606000, 2023YFA1606704; National Natural Science Foundation of China (NSFC) under Contracts Nos. 123B2077, 12035009, 11635010, 11735014, 11935015, 11935016, 11935018, 12025502, 12035013, 12061131003, 12192260, 12192261, 12342502, 12192262, 12192263, 12192264, 12192265, 12221005, 12225509, 12235017, 12361141819; the Chinese Academy of Sciences (CAS) Large-Scale Scientific Facility Program; the CAS Center for Excellence in Particle Physics (CCEPP); Joint Large-Scale Scientific Facility Funds of the NSFC and CAS under Contract No. U1832207, U2032104; CAS under Contract No. YSBR-101; The Excellent Youth Foundation of Henan Scientific Commitee under Contract No. 242300421044; 100 Talents Program of CAS; The Institute of Nuclear and Particle Physics (INPAC) and Shanghai Key Laboratory for Particle Physics and Cosmology; Agencia Nacional de Investigación y Desarrollo de Chile (ANID), Chile under Contract No. ANID PIA/APOYO AFB230003; German Research Foundation DFG under Contract No. FOR5327; Istituto Nazionale di Fisica Nucleare, Italy; Knut and Alice Wallenberg Foundation under Contracts Nos. 2021.0174, 2021.0299; Ministry of Development of Turkey under Contract No. DPT2006K-120470; National Research Foundation of Korea under Contract No. NRF-2022R1A2C1092335; National Science and Technology fund of Mongolia; National Science Research and Innovation Fund (NSRF) via the Program Management Unit for Human Resources \& Institutional Development, Research and Innovation of Thailand under Contract No. B50G670107; Polish National Science Centre under Contract No. 2019/35/O/ST2/02907; Swedish Research Council under Contract No. 2019.04595; The Swedish Foundation for International Cooperation in Research and Higher Education under Contract No. CH2018-7756; U. S. Department of Energy under Contract No. DE-FG02-05ER41374.

\end{document}